\documentclass[11pt,twoside]{article}
\usepackage{./asp2014}

\aspSuppressVolSlug
\resetcounters

\begin{document}

\title{Dwarf satellite galaxies in nearby groups of galaxies}
\author{Jaan~Vennik,$^1$ and Ulrich~Hopp,$^{2,3}$ 
\affil{$^1$Tartu Observatory, T\~oravere, Tartumaa, Estonia; \email{vennik@to.ee}}
\affil{$^2$Universit\"ats--Sternwarte M\"unchen, M\"unchen, Germany; \email{hopp@usm.lmu.de}}
\affil{$^3$MPI f. Extraterrestische Physik, Garching, Germany; \email{hopp@mpe.mpg.de}}}

\paperauthor{Jaan~Vennik}{vennik@to.ee}{}{Tartu Observatory}{}{T\~oravere}{Tartumaa}{61602}{Estonia}
\paperauthor{Ulrich~Hopp}{hopp@usm.lmu.de}{}{Universit\"ats--Sternwarte M\"unchen}{}{M\"unchen}{}{D 81679}{Germany}

\begin{abstract}
We analyse distribution, kinematics and star-formation (SF) properties of satellite galaxies in
three different samples of nearby groups. We find that studied groups are generally well approximated 
by low-concentration NFW model, show a variety of LOS velocity dispersion profiles and signs of SF quenching in 
outskirts of dwarf satellite galaxies.
\end{abstract}

\section{Motivation}
Within the CDM paradigm, dwarf satellite galaxies are expected to be particularly
sensitive to environmental effects and can serve as test particles when
studying the mass assembly history of dark matter halos.
In loose groups of low-concentration and with relatively shallow DM potential,
the accreting galaxies in the group infall region are subjected to
evolutionary mechanisms over larger time and spatial scales then in denser
systems. This may help to disentangle processes otherwise superimposed
in rich clusters. 
Here we analyse distribution, kinematics and star-forming (SF) properties of
satellite galaxies in five groups, observed by us with the Hobby-Eberly
Telescope (HET), and compare them with two larger samples of relatively
nearby groups.

\section{The target groups and the data}
Our main target sample consists of five reasonably isolated nearby groups,
centered on NGC 697, NGC 5005/33, NGC 6278, NGC 6962 and IC 65 (hereafter
HET groups). New dwarf members of these groups were found by our
spectroscopic HET observations and the new data are summarized in
Hopp \& Vennik (2014). For comparison, we make use of two larger samples
of nearby groups. The first comparison sample consists of 215   
medium-richness ($5 \le Ngal \le 40$), relatively nearby ($1000<cz<5000$ km/s)
groups selected from the parent catalog of SDSS DR10 (Tempel et al. 2014;
hereafter SDSS DR10 groups).
The second reference sample consists of 208 nearby ($cz < 3500$ km/s)
groups with at least 4 members in the northern ($\delta > 0^o$) part of the Local Supercluster (LScl)
selected from the catalog of Makarov \& Karachentsev (2011; hereafter LScl groups).
The photometric data in five optical ($ugriz$) and two GALEX UV-bands
($FUV, NUV$) as well as additional redshifts and other spectral data
were obtained from the NASA-Sloan Atlas.\footnote{\url{http://nsatlas.org}} 

\section{Structure and kinematics}

\articlefiguretwo{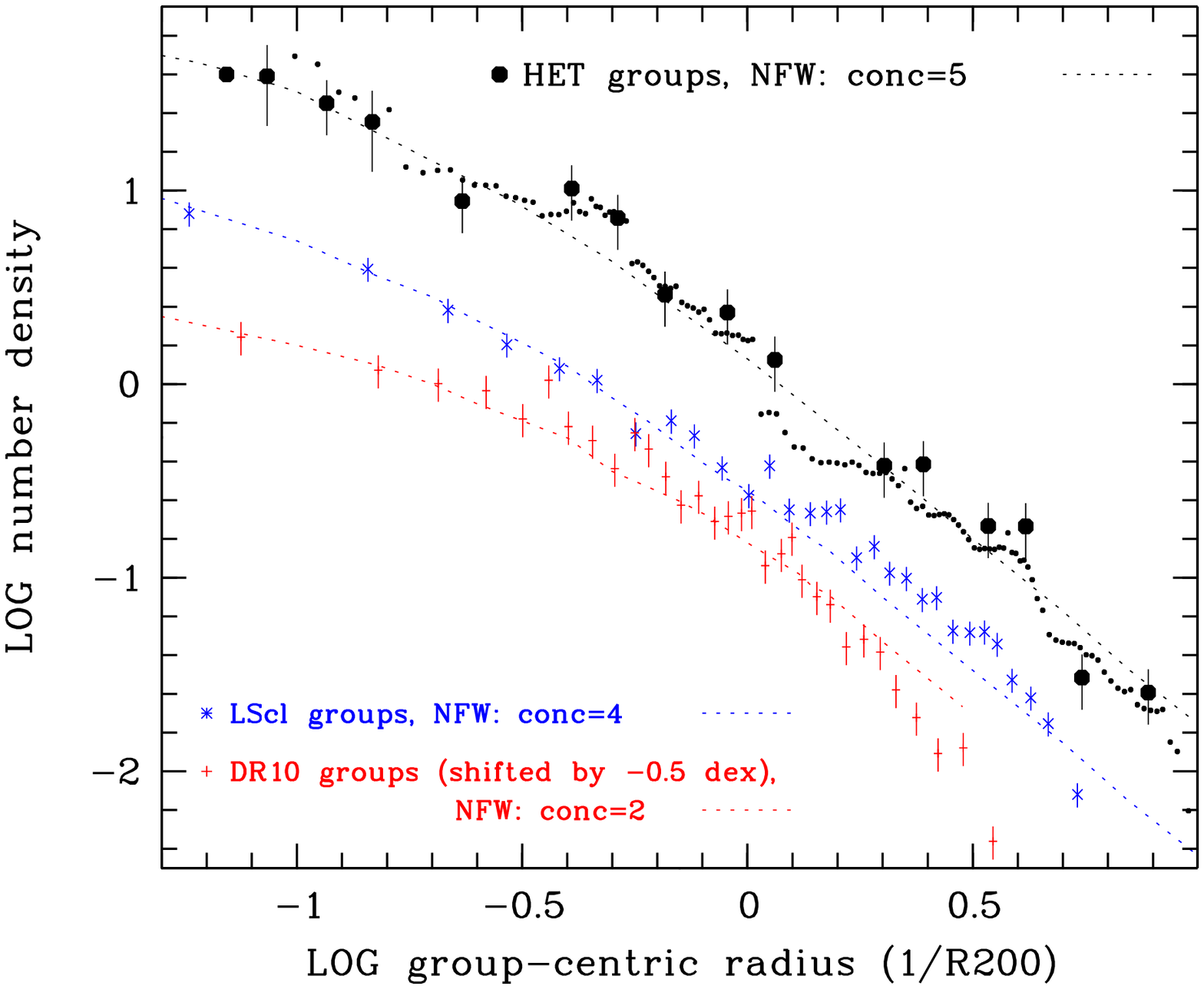}{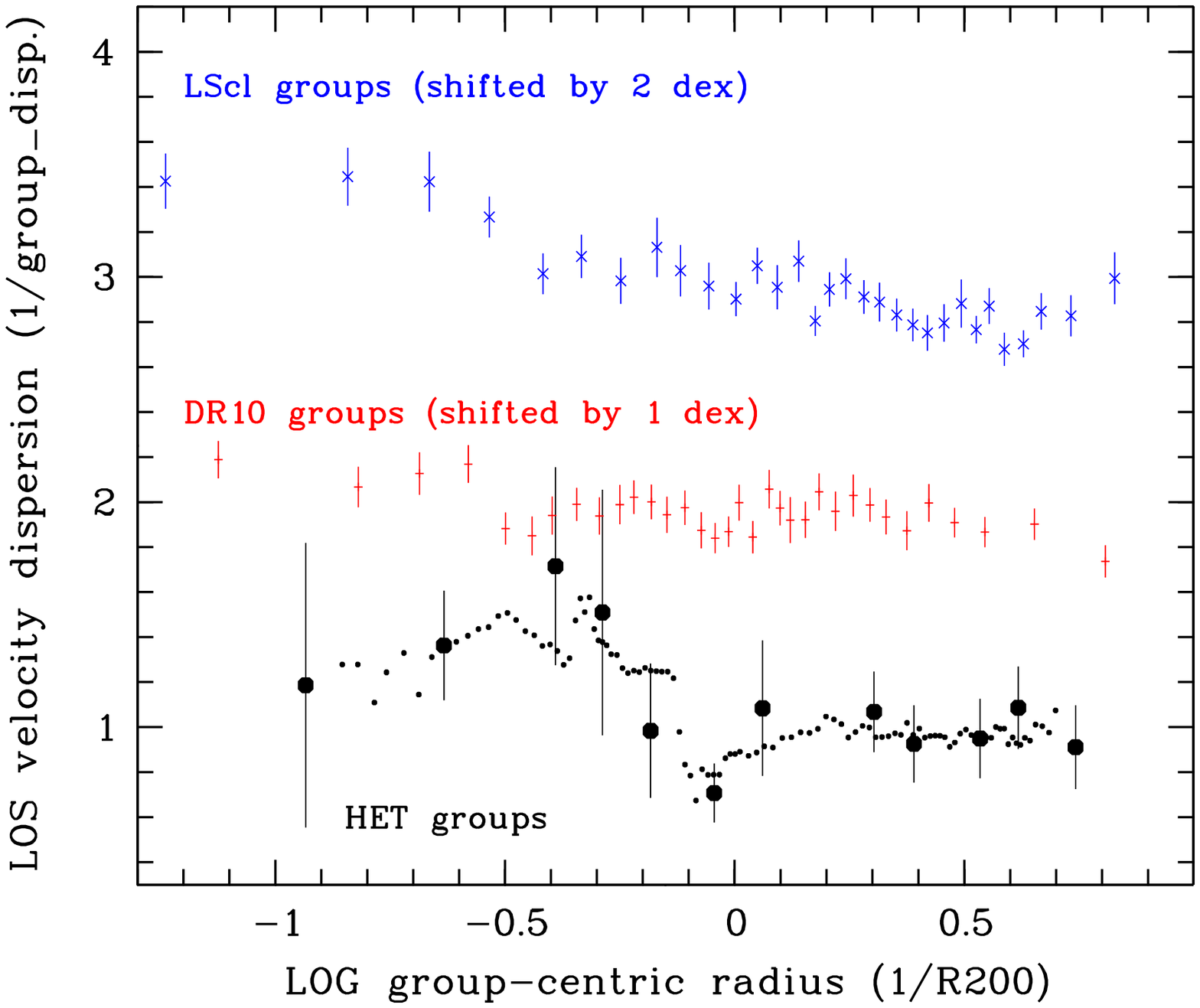}
{ex_fig1}{
\emph{Left:} Radial number density profiles (NDPs) of satellite galaxies in three composite groups.The NDP of the HET groups (top)
is computed twice: with bin size of 10 (near center 5) galaxies
(big dots, independent points) and NDP of moving group of 20 galaxies
(small dots). The NDPs of the LScl and DR10 (shifted by -0.5 dex) groups
are computed with bin size of 50 galaxies.
Error bars of the independent points are poissonian.
The presented NDPs are reasonably well fitted with low-concentration
($c = 2 - 5$) NFW models.
\emph{Right:} LOS velocity dispersion profiles (VDPs) are computed in the
same manner as the NDPs. Peculiar velocities of satellites are reduced by
the LOS velocity dispersion within $R_{200}$ of each LScl and DR10 group,
and by mean dispersion of HET groups ($<\sigma_v>=180$ km/s).
Errors of the independent points are estimated by jack-knife method.
}

We have computed the number-density profile of 152 satellite galaxies in
the HET groups,
as a function of reduced group-centric distance $R/R_{200}$
(where $R_{200} = \sqrt3\sigma_{v,200}/(10H_0)$).
Despite of poor statistics the radial density profile of composite group
is - in its full extent ($0.1 \le R/R_{200} \le 6$) -
reasonably well approximated by low-concentration ($c = 5\pm$1)
Navarro-Frenk-White (NFW) model (Fig.~\ref{ex_fig1}, left). 
Accordingly, the number-density profile of 1575 satellite galaxies in
the LScl groups 
is also well fitted by the NFW model of similar concentration ($c = 4 \pm 1$),
except some systematic departure in the periphery of composite group.
The second comparison sample of SDSS DR10 groups
is, within $\sim 1R_{200}$, reasonably well traced by 
a very low concentration ($c \simeq 2$) NFW model while it is steeply decreasing in the group' periphery.
Systematic deviation in periphery may manifest differences in the distribution of luminous (baryonic) and dark matter,
or result from poorly sampled outskirts of the DR10 groups.

The velocity dispersion profile (hereafter VDP) of the stacked HET group (Fig.~\ref{ex_fig1}, 
right, bottom) is noisy, however, 
it generally traces a trumpet-shape pattern, which has been observed from a number of rich clusters (e.g. Mahdavi et al.1999).
Satellite galaxies within $\sim 1R_{200}$ have a large velocity scatter, at larger radii of $\sim 1 - 5 R_{200}$
they manifest velocity characteristics of first infall. A density drop is evident also in the NDP of the stacked
HET group at nearly the same radius ($\sim 1R_{200}$), which
separates the inner, quasi-virialized and outer, infall domains.
The VDPs of two comparison samples 
exhibit flat cores and are more (LScl groups) or less (SDSS DR10 groups)
rapidly decreasing with group-centric radius.  
Stacking the data (positions and velocities) of many groups with a pronounced spread in concentration and
velocity distribution characteristics flattens both the mean NDPs and VDPs and smoothes out some
features (discontinuities), which could be useful e.g. for the group definition (Tully 2015).

\section{Star-forming properties}

\articlefiguretwo{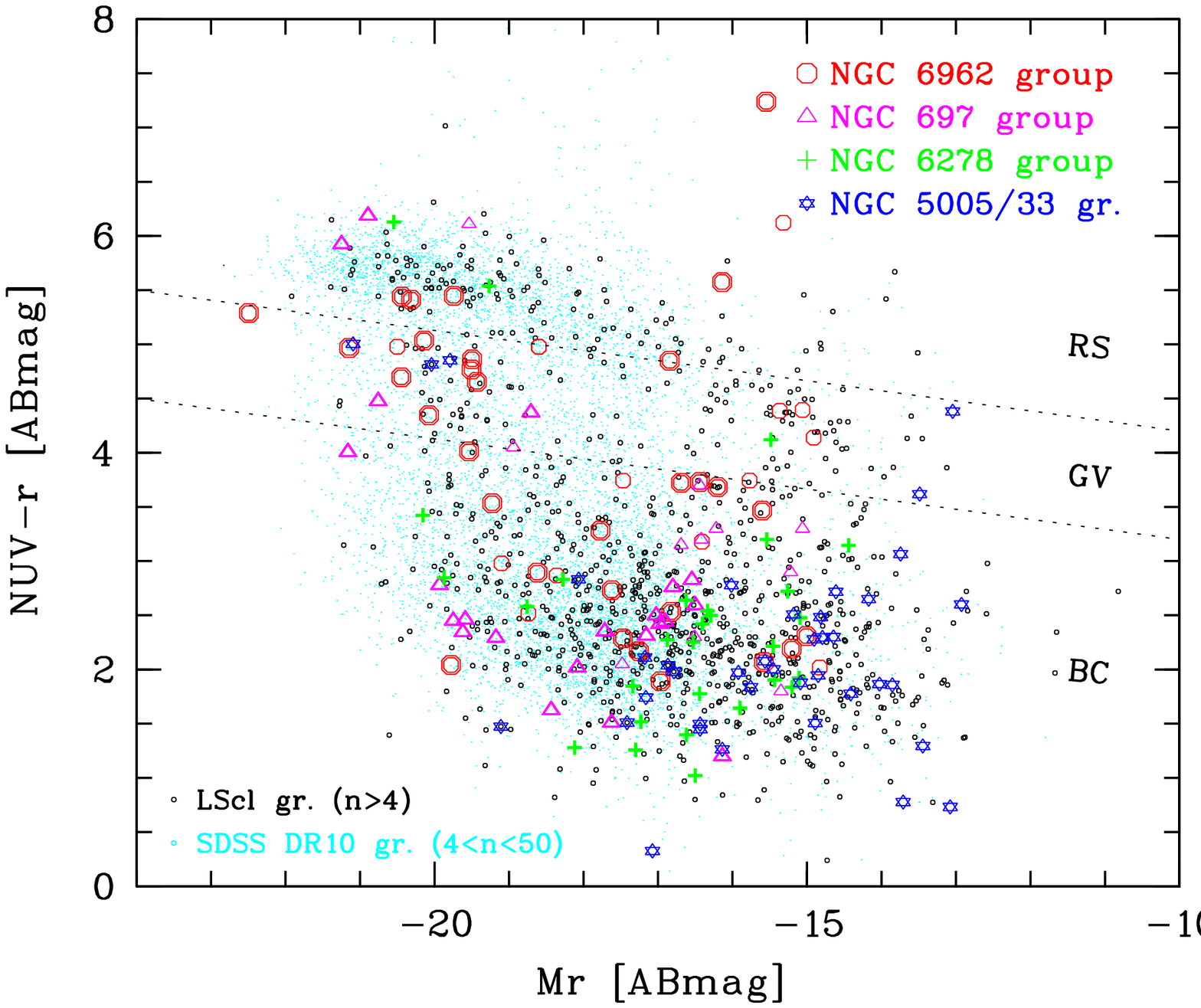}{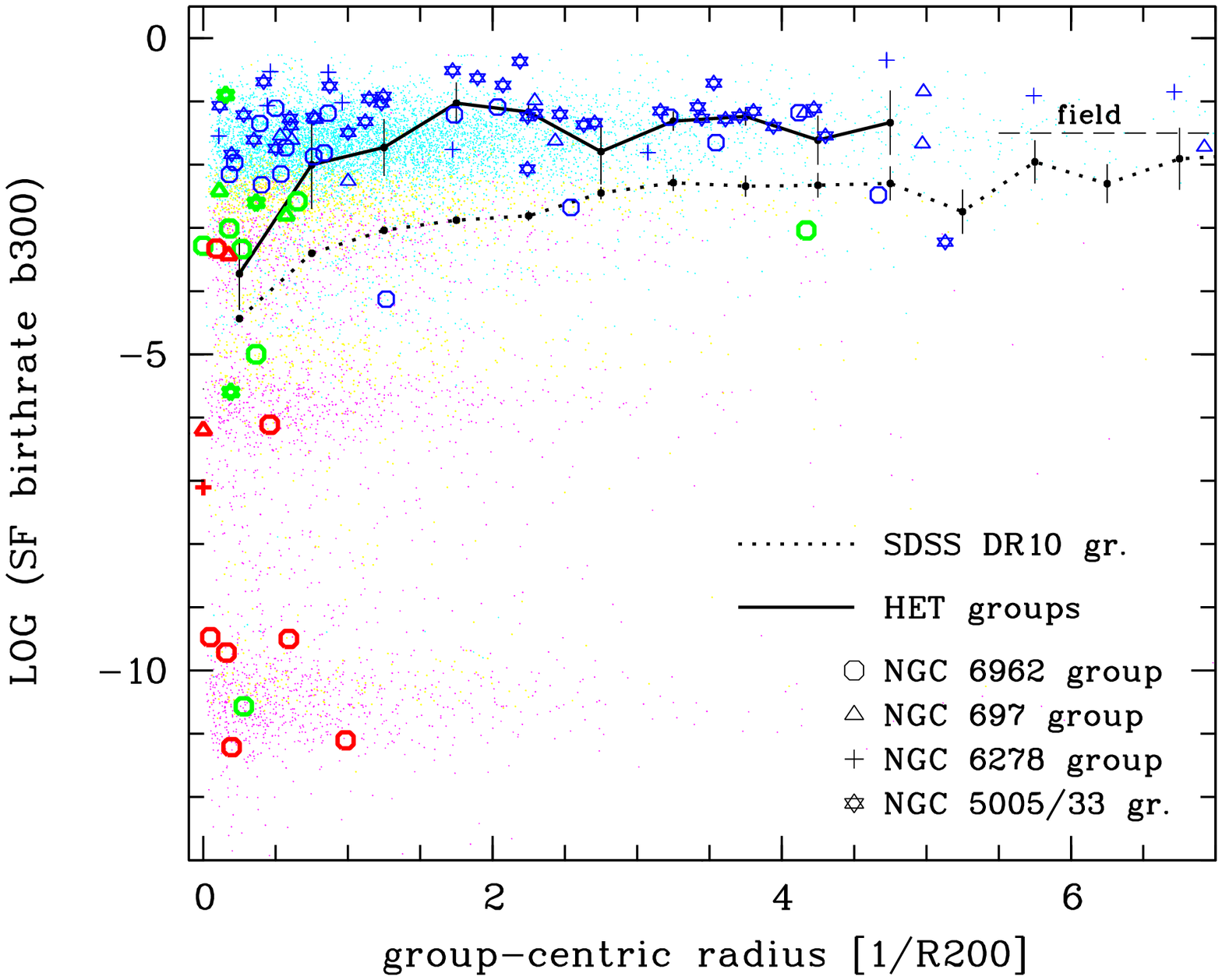}
{ex_fig2}{
\emph{Left:} The UV-optical colour-magnitude diagram of satellite galaxies in the SDSS DR10 groups (cyan),
in the LScl groups (black) and in four HET groups. The green valley (GV) of transitional galaxy candidates is arbitrarily
defined as  $2.1-0.1\times Mr < NUV-r< 3.1-0.1\times Mr$.
\emph{Right:} The distribution of the SF birth-rate parameter $b300$ of the RS, GV and BC satellite galaxies in the SDSS DR10 groups
(magneta/yellow/cyan dots, respectively), and in the four studied HET groups (red/green/blue symbols, respectively),
as a function of reduced group-centric distance.
}

The UV-optical color-magnitude diagram (CMD) is often used to
classify galaxies into three main groups: the blue cloud (BC)
of star-forming galaxies, the red sequence (RS) of passive/quenched
galaxies, and a third group of transitional galaxies, populating
the green valley (GV).        
The galaxies in the GV typically show low but non-zero SF
with specific SFRs in the range of $-11.8 < \log(SSFR) < -10.8$.
UV-optical colours, e.g. ($NUV-r$), are sensitive to low-level recent SF
and are thus good stellar age indicator, avoiding optical age-metallicity
degeneracy. 
Transition from blue to red galaxies may be driven by environmental processes,
associated with the infall into the group halo.
The distribution of the GV galaxies in dependence of group-centric radius
potentially allows to check various scenarios of dynamical evolution of
satellites moving inside the group massive haloes.\\
Candidates of transitional satellite galaxies were first selected in the
UV-optical CMD (Fig.~\ref{ex_fig2}, left). 
We used the spectral data from the NASA-Sloan Atlas, 
if available, to compare the luminosity-weighted ($NUV-r$) colour selection
with spectroscopic separation of star-forming and recently quenched galaxies,
based on specific SFR estimates over 300 Myrs timescale ($b300$).

The evolutionary stage of the galaxies in the studied groups differs:
members of the NGC 6962 and NGC 697 groups dominate the red sequence
and the green valley; most of the satellites of the NGC 5005/33 and NGC 6278
groups reside in the blue cloud, i.e. have younger stellar ages (Fig.~\ref{ex_fig2}, left) 
Remarkably, RS and GV satellites reside within the group virial radius while
BC galaxies spread far out (Fig.~\ref{ex_fig2}, right). The overall mean of the SF birth-rate $b300$ is
declining with diminishing radius within $\sim 2R_{200}$ (i.e. within the
group infall region); but the sudden drop in $b300$ within $\sim 1R_{200}$ is mostly due to recently
quenched satellites of the NGC 6962 group. 

\section{Summary}

\begin{itemize}
\checklistitemize
\item We find, consistent with earlier studies, the concentration of satellites
in loose groups ($c \simeq 2 - 5$) being
roughly two times smaller than typically predicted for DM haloes in
cosmological N-body simulations.
The offset between DM and satellite concentrations could probably result
from tidal evolution and merging processes
within $\sim 0.5 R_{vir}$ (Chen et al. 2006).
The concentration of DM halos is related to their formation redshift.
Low concentration groups and clusters are preferentially formed late
and are currently in accretion phase. 
\item Individual groups show a variety of LOS velocity-dispersion profiles (VDPs).
Declining VDP, as observed for the NGC 6962 group and for the two comparison
samples, qualitatively match N-body simulations of relaxed systems.
Other systems with irregular or rising VDP and with spatial structure
described by NFW profile are probably bound configurations but
not yet in dynamical equilibrium (Mahdavi et al. 1999). 
\item The luminosity-weighted colour classification of galaxies into RS, GV and BC,
generally, traces the distribution of the birth rate
parameter ($b300$), as determined from fiber-spectroscopy. The spread
of RS and GV galaxies towards high $b300$ values probably results from
low-level nuclear SF. This is in-line with recent observation, that the
GV galaxies with stellar mass $<10^{10} M_\odot$ tend to show flat or rising
(i.e. getting redder with increasing radius) colour gradient - indicative of
outside-in SF (Pan et al. 2015).  A similar colour trend was found for
dwarf galaxies of the HET groups (Hopp \& Vennik 2014).\\
\item We conclude that strangulation could be the primary mechanism for quenching SF in outskirts 
of (dwarf) satellite galaxies, with timescales of $\sim 4$ Gyrs (Peng et al. 2015),
possibly at work in hot haloes of groups and clusters.
\end{itemize}

\end{document}